\documentclass[sn-basic]{sn-jnl}

\usepackage{graphicx}%
\usepackage{multirow}%
\usepackage{amsmath,amssymb,amsfonts}%
\usepackage{amsthm}%
\usepackage{mathrsfs}%
\usepackage[title]{appendix}%
\usepackage{xcolor}%
\usepackage{textcomp}%
\usepackage{manyfoot}%
\usepackage{booktabs}%
\usepackage{algorithm}%
\usepackage{algorithmicx}%
\usepackage{algpseudocode}%
\usepackage{listings}%

\theoremstyle{thmstyleone}%
\theoremstyle{thmstyletwo}%

\theoremstyle{thmstylethree}%

\begin{document}

\title[Article Title]{Formation of classical Be-stars of the early spectral subclass in the case of nonconservative mass transfer in close binary systems}


\author*[1]{\fnm{Evgeny} \sur{Staritsin}}\email{evgeny.staritsin@urfu.ru}



\affil*[1]{\orgdiv{Astronomical Observatory}, \orgname{B.N. Yeltsin Ural Federal University}, \orgaddress{\street{Mira-19}, \city{Ekaterinburg}, \postcode{620062}, \country{Russia}}}




\abstract{Spin-up of a mass gaining component in a binary system is considered taking into account the mass loss from the system during the mass transfer between components in the Hertzsprung gap. The angular momentum that the accreting component gains during mass transfer depends on the increase in the mass of the component at this stage. The increase in the mass was considered over a broad range, from 5\% to 100\%. The case is considered when, after mass transfer, the mass of the accreting component has a value of 16~$M_\odot$, typical for early Be stars. The transfer of angular momentum within the accreting component occurs due to meridional circulation and shear turbulence. If the accreted mass accounts for more than 30\%, the accretor obtains a rotation typical of early Be-stars. This conclusion does not depend on: a) the rotation of the accreting component before mass transfer, b) the amount of angular momentum coming from the boundary layer located between the star and the accretion disk, c) a possible decrease in the angular velocity of the disk below the Keplerian value, d) the efficiency of turbulence in the interior of the accretor.}




\maketitle

\section{Introduction}\label{sec1}

Classical Be stars manifest themselves through emission in Balmer lines \citep{pr03}. These stars are at the stage of hydrogen burning in their central parts. Observations generally show that Be stars rotate faster than normal B-type stars \citep{c2005, hgs10, dld13}. Be stars of the early spectral subclass have rotation velocities greater than 40-60\% of the Keplerian velocity \citep{c2005}. The origin of such rotational velocities is still the subject of modern research.

Rapid rotation of Be stars may be due to an interaction with a nearby star, with which the Be star forms a close binary system \citep{pcwh91, pz95, sl14, sl21, hlws21}. The more massive star of the binary system, the donor, increases its size due to evolution, fills the Roche lobe, and begins to lose mass. Some of this lost mass falls onto the other star in the system, called the accretor, bringing with it a significant amount of angular momentum. Therefore, the mass and angular momentum of the accretor increase. Another part of the lost mass leaves the binary system, carrying a certain amount of angular momentum. The mass lost from the system is determined by a number of parameters, such as the evolutionary status of the components, the ratio of the components masses, the mechanism of energy transfer in the components envelopes, and so on \citep{mt1988}. A few attempts have been made to determine the parameters of the non-conservative mass transfer process in Algols, taking into account the hot spot properties and the possible presence of magnetic fields \citep{dti2010, rgm11, dsd13, rg2020}. 
Various parameterizations of mass and angular momentum loss from a binary system are used to explain the observed properties of Be stars in population synthesis studies \citep{ty79, pcwh91, pz95, bv97, 2020MNRAS.498.4705V, sl21, rkd2024}. The issue of non-conservative mass transfer in binary star systems remains a complex and unsolved problem to this day.

The direct evidence for non-conservative mass exchange in binary systems is very limited and difficult to study. Perhaps FS CMa type stars provide such evidence. These stars show strong emission in forbidden and permitted optical lines, as well as infrared excesses \citep{as76}. The interpretation of these observations suggests that FS CMa stars are binary systems undergoing non-conservative mass transfer \citep{mli13}.  The presence of circumstellar matter makes it difficult to determine the exact masses of the individual components. In addition, measuring the amount of matter surrounding the system is difficult.

The population synthesis method allows us to estimate the amount of mass that should fall on the accretor and the amount of mass that can be lost from the system to explain the observed properties of the population of Be stars. The periods of Galactic Be binaries with a helium star companion and the masses of Be stars in these binaries can be explained by a conservative mass exchange in the Hertzsprung gap \citep{sl21}. The lower limit of the mass of Be stars in X-ray binaries is 6-8  $M_\odot$. This limit can be explained in the case of effective accretion, when the accretor acquires more than 30\% of the mass lost by the donor \citep{2020MNRAS.498.4705V}. The range of periods and eccentricities of X-ray binaries with Be stars is reproduced when the increase in the mass of the Be star is 5-10\% \citep{rkd2024}. Comparisons between population-synthesis predictions and observations suggest a wide range of accretion efficiencies \citep{2020MNRAS.498.4705V, sl21, rkd2024}.

Binary systems provide us with the most accurate knowledge of the masses and radii of stars. In particular, the stellar parameters of the binary system $\phi$~Per, consisting of a Be-star and an O-subdwarf, have been reliably determined \citep{gbfk98, sgmg18}. In the past, the $\phi$~Per system has experienced the first stage of mass transfer. The current parameters of this system are predicted under the assumption of a conservative mass exchange in the Hertzsprung gap, and the mass of the Be-star should have almost doubled \citep{vlr98, sgmg18}. An increase in the mass of the accreting component by a few percent leads to a Keplerian break up rate \citep{p1981}. 
However, as argued by \cite{p1991} and \cite{pn91}, accretion from the disk does not necessarily cease when the break up rate is reached. The accretion disk regulates the mass and angular momentum flux through viscous coupling, allowing the star to continue accreting. Specifically, there are both accretion and decretion solutions for the same spin rate. However, if there is infall accretion on to the disk away from the inner boundary, this may drive accretion \citep{mlv25}. In the case of mass transfer in a binary system, matter from the star filling the Roche lobe flows into an accretion disk surrounding the second star. As long as the inflow of matter into the disk continues, the second star can accrete matter even if its rotation is critical \cite{p1991}. The increase in the accretor's mass is accompanied by an increase in its angular momentum along the sequence of critically rotating states \citep{bk93}. This increase in angular momentum is less than the Keplerian one. Excess angular momentum is transported from the accretor to the outer boundary of the accretion disk due to viscosity. 
Consequently, the mass and angular momentum of the accreting component can grow while the accreting component itself remains in a state of critical rotation \citep{p1991, bk93, mlv25}.

The transfer of angular momentum within the accreting component was considered by \cite{St2022, St2024a}. The processes of angular momentum transfer are meridional circulation and turbulence. 
In the critical rotation state of the accreting component, the meridional circulation transfers most of the angular momentum of the accreted mass to the surface of the accreting component. This part of the angular momentum enters the disk and can then be transferred through the disk further away from the accreting component according to \cite{ p1991} and \cite{bk93}. The angular momentum remaining in the accreted mass is sufficient for the accreting component to have a rotation similar to that of the Be-star in the binary system $\phi$~Per \citep{St2024b}.

In general, mass transfer in a binary system can be non-conservative. Let's consider a series of binary systems with different component masses. Let's assume these systems have completed the mass transfer stage. We'll select those systems whose accretors have the same mass. The initial mass of these accretors was different in each system. Consequently, each accretor has received its own mass increment. Along with the mass, the star has acquired angular momentum. Thus, the angular momentum of the accretor depends on the fraction of the accreted mass. How much mass does the accreting component have to gain for the resulting rotation to be the same as that of the Be-star? In this paper, we consider the increase in angular momentum of the accreting component in dependence on the increase in its mass. To specify the mass that the accretor gained due to mass transfer, we choose 16~$M_\odot$, since this value is typical for stars of spectral subclass B0-B2.

Thus, we accept the following scheme of the spinning up of the accreting component during mass transfer in a binary system. Part of the mass lost by the star filling the Roche lobe leaves the binary system, while the rest falls onto the second star - the accretor. The accretor's mass increases continuously throughout the entire mass transfer phase of the binary system. We calculate the structure and evolution of the accreting component in detail and account for the transfer of angular momentum in its interior in accordance with the theory of \cite{Zahn92}. The angular momentum exchange between the accreting component and the accretion disk is accounted for through boundary conditions. As a result, the angular momentum acquired by the accreting component is determined as a function of its mass gain during the mass transfer stage in the binary system.

\section{Basic equations and simplifications}\label{sct2}

We consider the structure and evolution of the accreting component during the mass transfer in a binary system, and the angular momentum exchange between the accreted mass and the inner layers of the accretor is taken into account. The evolution of the accreting component is calculated in a one-dimensional approximation. The evolutionary code created by \cite{p1970} was modified to account for rotation \citep{St99}. The independent variable is the mass of matter inside a constant pressure surface. The influence of rotation on the structure of the component is taken into account according to Clairaut’s theory \citep{t1978}. Models of rotating isolated main sequence stars show a small angular velocity difference between the convective core and the surface \citep{mm05, St07, emmb08, geg13}. Therefore, even in the case of critical rotation, the compression of the constant pressure surfaces in the inner region, which contains 99\% of the mass of the star, does not exceed 5\% \citep{St07}. 
Small values of the difference in angular velocities between the convective core and the surface, obtained in the one-dimensional approximation, are confirmed both by two-dimensional modeling \citep{er13} and by the analysis of the variable star properties using the asteroseismology method \citep{bbm23}.

\subsection{Transfer of angular momentum inside of the accretor}\label{ssct3}

The angular velocity in the radiative envelope can vary in the vertical direction, with the angular velocity almost constant along the isobar \citep{Zahn92}. We have considered the transfer of angular momentum in the radiative envelope of the accreting component. This transport occurs because of the circulation of matter in the meridional plane and because of shear turbulence. The transfer is described by the following equation \citep{t1978}:
\begin{equation}
\label{eq003}
\frac{\partial(\rho\varpi^2\Omega)}{\partial t}+
\mbox{div}(\rho\varpi^2\Omega{\bf u})
=\mbox{div}(\rho\nu_{\mbox{v}}\varpi^2\mbox{grad}\Omega).  
\end{equation}
The circulation rate in the meridional plane $\bf u$ is determined according to \cite{mz98}:
\begin{equation}
\label{eq004}
\rho T{\bf u}\mbox{grad}s=\rho\varepsilon_n+
\mbox{div}(\chi\mbox{grad}T)-\mbox{div}{\bf F}_h.  
\end{equation}
In these equations,
$\Omega$ - angular velocity,
$\varpi$ - distance to the rotation axis,
$\rho$ - density,
$\nu_{\mbox{v}}$ - turbulent viscosity in the vertical direction,
$T$ - temperature,
$s$ - specific entropy,
$\varepsilon_n$ - nuclear energy release rate,
$\chi$ - thermal conductivity,
${\bf F}_h$ - turbulent enthalpy flux in the horizontal direction:
${\bf F}_h=-\nu_h\rho T\partial{s}/\partial{\bf i_\theta}$
and $\nu_h$- turbulent viscosity in the horizontal direction.
The rotation of the convective core is assumed to be solid.

\cite{es78} proposed a diffusion approximation to describe the transport of angular momentum in the interior of a rotating star. This simplification involves the transfer of angular momentum in one direction: from the region of rapid rotation to the region of slow rotation. The diffusion approximation was further developed in the works of \cite{pks89} and \cite{hlw00}, as well as taking into account the magnetic field \citep{sp02, hws05}. The same approximation is used to study the rotation in the interior of accreting components of binary systems \citep{plh05, wls20}. However, equation (\ref{eq003}) cannot be reduced to a diffusion form \citep{cz92}. We consider the transfer of angular momentum on the basis of equations (\ref{eq003}) and (\ref{eq004}). To solve these equations, the first order of smallness is taken into account in the expansion of the vertical component of the meridional circulation velocity in latitude: 
\begin{equation}
\label{eqU}
U_V(r,\theta)=U(r)\cdot\mbox{P}_2(cos\,\theta). 
\end{equation}
Here, $U(r)$  - amplitude of the vertical component of the meridional circulation velocity, $\mbox{P}_2(cos\,\theta)$ - the Legendre function of order 2. Such an approximation makes it possible to reduce the equations to one-dimensional ones \citep{Zahn92}. These equations are solved together with the equations for structure and evolution of the accreting component \citep{St99, St05, St07, St19a}. As a result, the structure of the accreting component, the angular velocity $\Omega(r)$, and the amplitude of the vertical component of the circulation $U(r)$ are determined, as well as the evolution of these parameters over time.

Once $U(r,t)$ and $\Omega(r,t)$ are determined, it is possible to calculate the flux of angular momentum carried by circulation (advective flow) \citep{Zahn94}:
\begin{equation}
F_{ad}=-\frac{8\pi}{15}r^4\Omega\rho U, \nonumber
\end{equation}%
as well as the turbulent flux of angular momentum:
\begin{equation}
F_{t}=-\frac{8\pi}{3}\nu_{V}r^4\rho\frac{d\Omega}{dr}, \nonumber
\end{equation}%
and the total flux:
\begin{equation}
F=F_{ad}+F_{t}. \nonumber
\end{equation}%

Turbulent viscosity in the vertical and horizontal directions is determined according to \cite{Zahn92}, \cite{TZ97}, \cite{M2003}, \cite{MPZ04}:
\begin{equation}
\label{Vis_V}
\nu_V=\frac{2Ri}{N_T^2/(K+D_h)+N_\mu^2/D_h}
\left[\frac{d(\varpi\Omega)}{dr}\right]^2, 
\end{equation}

\begin{equation}
\label{Vis_h}
\nu_h=C\cdot r\cdot\left|U(r)\right|. 
\end{equation}
Here $Ri$ - the critical Richardson number,
$N^2=N_T^2+N_\mu^2$,
$N$ - buoyancy frequency,
$K$ - thermal diffusivity.
The applicability condition of the Zahn model \citep{Zahn92} is as follows $\nu_V<\nu_h<K$.

The main uncertainty in Zahn's model is the efficiency of turbulent angular momentum transfer. To assess the role of turbulence, the spinning up of the accreting component was analyzed with artificially suppressed turbulence \citep{St2022}, as well as with turbulence taken into account \citep{St2024a}, with turbulent viscosity determined using (\ref{Vis_V}) and (\ref{Vis_h}). Continuing to elucidate the significance of turbulent transfer, we consider here the most efficient turbulence. For this, the turbulent viscosity in the horizontal direction is defined as $\nu_V=K$. This can be considered as the upper limit of the turbulence efficiency.

\subsection{Angular momentum input to the accretor, boundary conditions}\label{ssct2}

We suppose that some of the mass lost by the donor forms the Keplerian disk around the accretor \citep{ls75, rr1998, bhbk2000, r2012}. At the beginning of mass transfer, the angular velocity of the accretor is less than the Keplerian velocity, so a boundary layer forms between the accretor and the disk \citep{p1991, pn91, cnc91, pn95, hk17, dja21}. In this layer, the angular velocity decreases from the critical velocity in the disk to the value on the surface of the accretor. A decrease in angular velocity may occur due to the removal of angular momentum from the boundary layer.

The processes of angular momentum removal from the boundary layer are the subject of current research. A number of boundary layer models are constructed assuming turbulent transport of angular momentum toward the upper part of the star \citep{pn95, hk17, dja21}. 
Accretion disks tend to move supersonically near the stellar surface. This makes them resistant to Kelvin-Helmholtz instability \citep{m1958}, limiting the hydrodynamic mechanisms for driving turbulence. Supersonic shear instabilities \citep{br12} can generate waves capable of transporting angular momentum through the boundary layer \citep{brs12}. These waves can transfer angular momentum from the boundary layer to the upper part of the accretor and to the upper part of the disk \citep{trp25}. The properties of the boundary layer depend on the heating and cooling processes that operate in it \citep{2024ApJ...974..218D}. These processes have been poorly studied to date, and the full picture of what happens within the boundary layer is still unknown.

We have considered two opposite cases of angular momentum transfer from the boundary layer. In one case, the angular momentum is transferred from the boundary layer to the upper part of the accretor, which is an example of an effective boundary layer. The angular momentum of the accretor $J$ increases with the rate:
\begin{equation}\label{e1}
\frac{dJ}{dt}=\frac{2}{3}R^2(\Omega_{c}-\Omega_s)\dot M,
\end{equation}%
where $t$ is time, $\Omega_{c}$ is the critical velocity of the disk defined near the accretor surface, $\Omega_s$ is the angular velocity of the accretor surface and $\dot M$ is the mass accretion rate. Another case is an inefficient boundary layer – the angular momentum cannot be transferred from the boundary layer to the upper part of the accretor.

In any case, the mass added to the accretor has the same angular velocity as the surface of the accretor. The specific angular momentum has the greatest value on the surface of the accreting component. Adding a mass with this specific angular momentum increases the average angular momentum of the accreting component. As a result, the mass, angular momentum, and angular velocity of the surface of the accreting component increase \citep{St19a}. In the case of an effective boundary layer, the accreting component receives an additional amount of angular momentum. In this case, the boundary condition is determined by formula (\ref{e1}). If there is an ineffective boundary layer, the boundary condition has the form: $dJ/dt=0$.

Soon after accretion begins, the angular velocity of the surface of the accreting component becomes equal to the critical value. In this case, the angular velocity of the disk decreases monotonically with distance from the star. Therefore, the viscous torques transport angular momentum outward. The star can lose angular momentum, which is transferred out from the star to the disk \citep{p1991}. If there is no external influx of matter into the disk, accretion can give way to decretion, with a decrease in the star's angular velocity \citep{pn95, hk17, dja21, mlv25}. However, in the case of mass transfer in a binary system, matter continues to flow into the disk from the donor due to its overflowing Roche lobe. Assuming rigid-body rotation, the star reaches Keplerian break up rate after its mass increases by 5-10\% \citep{p1981}. Data on the masses and radii of binary systems consisting of a Be star and an OB subdwarf indicate the possibility of a nearly twofold increase in the mass of the Be star's progenitor during the mass transfer in the binary system \citep{sgmg18, lmv25}. The influx of matter into the disk establishes an accretion regime \citep{mlv25}. Therefore, we accept the hypothesis proposed by \cite{bk93} about accretion onto a star in a state of critical rotation.

Let us consider a sequence of stars in a state of critical rotation, with $J$ denoting the angular momentum of a star with a mass of $M$. Then $0<j_a=dJ/dM<j_{Kep}$, where $j_{Kep}$ is the specific Keplerian angular momentum at the equator of the star. This means that the angular momentum of a star in a state of critical rotation increases with an increase in its mass by an amount less than the Keplerian value. As shown by \cite{p1991} and \cite{bk93}, the excess angular momentum $\delta j=j_{Kep}-j_a$ can be moved across the disk farther away from the star. As a result, the mass and angular momentum of the star can increase, and the star remains in a state of critical rotation. Therefore, after the rotation of the accreting component becomes critical, the boundary condition has the form: $\Omega_s=\Omega_{c}$.

\subsection{Binary system initial parameters}\label{ssct1}

\begin{table}
\caption{ Parameters of calculated binaries.}\label{tbl}%
 \begin{tabular}{ccccccc}
\toprule
$\zeta$ & ${(M_d)_i+(M_a)_i}$ & $A_i$ & $\dot M$ & $(J_a)_i$ & $(J_a)_f$ & $(J_a)_{f}^\prime$ \\
 & [${M_\odot}$] & [${R_\odot}$] & [${M_\odot}\,year^{-1}$] & & [$g\cdot cm^2s^{-1}$] & \\
\midrule
0.05 & 32.0+15.2 & 67.0 & 0.56$\times10^{-4}$ & 3.1$\times10^{51}$ & 2.8$\times10^{52}$ &  1.1$\times10^{52}$ \\
0.10 & 29.8+14.4 & 67.2 & 1.06$\times10^{-4}$ & 2.7$\times10^{51}$ & 3.2$\times10^{52}$ &      2.8$\times10^{52}$ \\
0.15 & 27.6+13.6 & 67.5 & 1.46$\times10^{-4}$ & 2.3$\times10^{51}$ & 3.4$\times10^{52}$ &      3.2$\times10^{52}$ \\
0.20 & 25.3+12.8 & 67.9 & 1.77$\times10^{-4}$ & 1.8$\times10^{51}$ & 3.6$\times10^{52}$ &      3.5$\times10^{52}$ \\
0.30 & 20.9+11.2 & 68.7 & 2.13$\times10^{-4}$ & 1.2$\times10^{51}$ & 4.1$\times10^{52}$ &      3.9$\times10^{52}$ \\
0.40 & 16.4+9.6 &  70.0 & 2.16$\times10^{-4}$ & 6.7$\times10^{50}$ & 4.5$\times10^{52}$ &      4.4$\times10^{52}$ \\
0.50 & 12.0+8.0 &  72.0 & 2.08$\times10^{-4}$ & 3.7$\times10^{50}$ & 4.9$\times10^{52}$ &      4.8$\times10^{52}$ \\
\botrule
\end{tabular}
\footnotetext{The following is indicated in the columns:
$\zeta$ -- the fraction of mass accreted,
${(M_d)_i+(M_a)_i}$ -- the initial masses of the donor and accretor,
$A_i$ -- the initial distance between components,
$\dot M$ -- the average accretion rate,
$(J_a)_i$ -- the initial angular momentum of the accretor,
$(J_a)_f$ -- the final angular momentum of accretor for an effective boundary layer,
$(J_a)_{f}^\prime$ -- the final angular momentum of accretor for an ineffective boundary layer}
\end{table}

We have considered mass transfer in binary systems with different initial masses of the components. Mass transfer begins once the donor completes evolution in the main sequence. As a result of mass transfer in a binary system, the mass of the accretor increases by $\Delta M_a$. In all cases, it is assumed that the mass of the accretor after the end of the mass transfer $(M_a)_f$ is equal to 16 M$_\odot$. The initial mass of the accretor is determined according to $(M_a)_i=(M_a)_f-\Delta M_a$. We consider seven different values of $\zeta$, where $\zeta=\Delta M_a/(M_a)_f$ (Table \ref{tbl}). The mass transfer efficiency—the fraction of mass retained in the accretor during the mass transfer phase—remains uncertain. An estimate of the mass transfer efficiency is obtained from an analysis of a sample of 16 binary systems, each system consisting of a Be star and an OB subdwarf with well-constrained masses \citep{lmv25}. The given range of $\zeta$ is consistent with this estimate. The detailed theory of mass loss from a binary system is still under development \citep{dti2010, rgm11, dsd13,dbj15, rg2020}. To determine the initial mass of the donor $(M_d)_i$, we assume that at $\zeta$=0.05, the mass of the donor and the mass of the accretor are equal after completion of mass transfer, and at $\zeta$ = 0.5, the mass transfer is conservative. We used the relationship between the donor's final mass and the donor initial mass \citep{mt1988}:
\begin{equation}
(M_d)_f=\eta(M_d)_i^{1.4}, \nonumber
\end{equation}%
where $\eta=0.122$, as in \cite{St2024b}.
The initial masses of the components of the binary system are shown in the second column of the Table \ref{tbl}.

Since both conservative and non-conservative mass transfer in a binary system are considered, the initial distance between the components is chosen near the boundary separating these two modes of mass transfer. To determine the mode of mass transfer, we based on a simple relationship between the thermal time scales of the donor $(t_{KH})_d$ and the accretor $(t_{KH})_a$ \citep{kmh77}, where
\begin{equation}
t_{KH}=3.12\times10^7\frac{M^2}{RL}. \nonumber
\end{equation}%
In particular, if $(t_{KH})_a<C\times(t_{KH})_d$, mass transfer occurs conservatively; in contrast, if $(t_{KH})_a>C\times(t_{KH})_d$, the binary system is losing mass. $C=10$ is the generally accepted value \citep{htp02}. However, the parameter $C$ may depend on the initial accretor mass, according to \cite{lhm24}. If $(M_a)_i<5M_\odot$, then $C\ge100$. In the range $8M_\odot<(M_a)_i<20M_\odot$, the parameter $C\approx 10$. The value $C=10$ is used for the boundary separating the two mass transfer modes.

The ratio between the donor radius $R_d$ and the accretor radius $R_a$ for the boundary case $(t_{KH})_a=10\times(t_{KH})_d$ can be found using the mass-luminosity relation $L\sim M^3$:
\begin{equation}
R_d=10\cdot R_a\cdot M_a/M_d. \nonumber
\end{equation}%
Since we consider mass transfer in the Hertzsprung gap, the age of the binary system, when mass transfer begins, approximately equals the age of the donor on the main sequence. All parameters of the accretor structure for this age can be determined \citep{St18}. The distance between the components is determined by the formula:
\begin{equation}
A=R_d/(0.38+0.21\cdot\log(M_d/M_a)). \nonumber
\end{equation}%
This distance (Table \ref{tbl}) is assumed to be the initial distance. It is not necessary to know the exact value of the initial distance. The angular momentum of the accretor after completion of mass transfer in the Hertzsprung gap does not depend on small variations in the initial distance between the components, all other things being equal \citep{St2024b}.

The duration of mass transfer is approximately three times the thermal time scale of the donor \citep{p1971}. At the selected initial distance between the components, the mass transfer time is one third of the thermal time scale of the accretor. Because of non-conservative mass transfer, the rate of mass accretion by the accretor is less than the rate of mass loss by the donor. In simple consideration, we assume the mass accretion rate to be constant and equal to the average. This average rate of mass accretion is shown in the fourth column of Table \ref{tbl}.

The synchronization time of the axial rotation of the accretor with orbital motion exceeds the lifetime of the donor star on the main sequence for all systems listed in Table \ref{tbl} \citep{htp02, Zahn75}. Before mass transfer begins, the accretor may have any axial rotation. We have considered the case of synchronous rotation of the accretor. The initial angular momentum of the accretor is indicated in the fifth column of Table \ref{tbl}.

An increase in the mass of the accreting component causes the growth of its convective core. Because of mixing, the hydrogen content in the convective core and the overlying layer changes. We consider the case where the distribution of hydrogen in the accreting component is the same as that of the isolated star, and the mass and total helium content of the isolated star are similar to those of the accreting component \citep{H1983}.

\begin{figure}[h]
\centering
\includegraphics[width=0.9\textwidth]{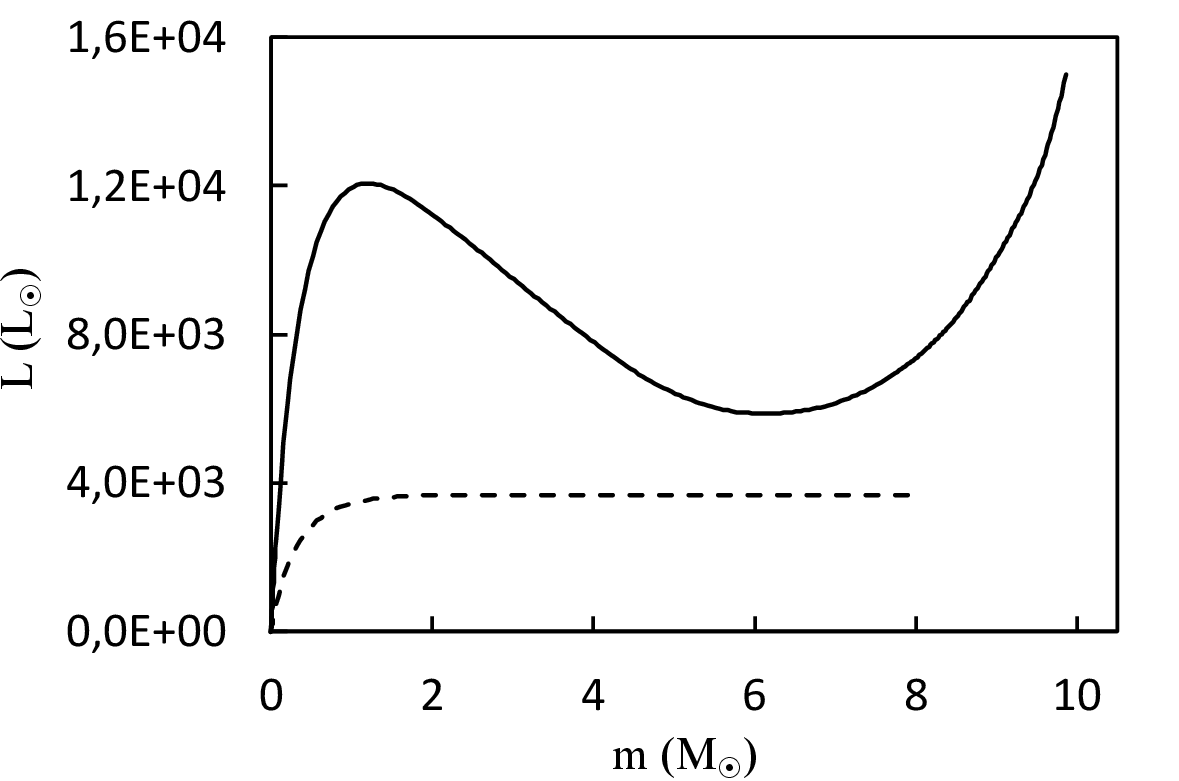}
\caption{Luminosity profile in the interior of the accreting component with an initial mass of 8~$M_\odot$ before mass transfer (dashed line) and after an increase in mass by 2~$M_\odot$ (solid line).}\label{fig01}
\end{figure}

\begin{figure}[h]
\centering
\includegraphics[width=0.9\textwidth]{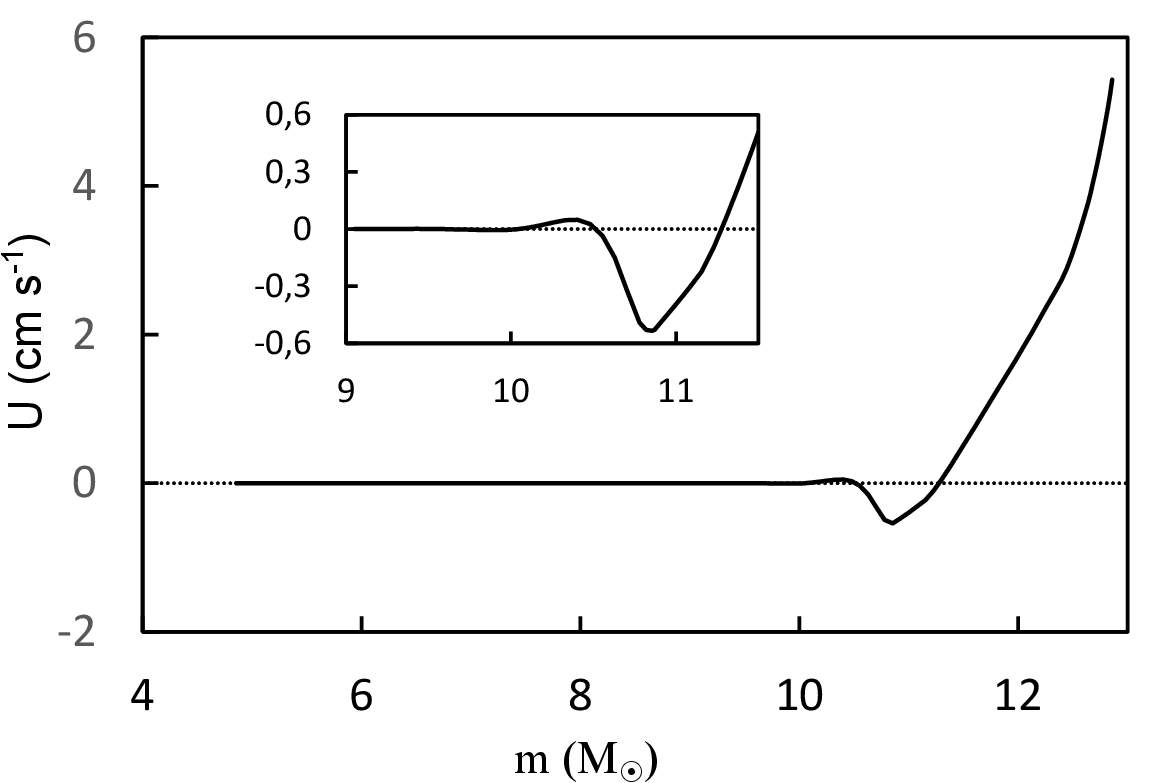}
\caption{The amplitude of the vertical component of the meridional circulation velocity inside an accretor with an initial mass of 12.8~$M_\odot$ after a mass increase of 0.24~$M_\odot$ (solid line).}\label{add0101}
\end{figure}

\begin{figure}[h]
\centering
\includegraphics[width=0.9\textwidth]{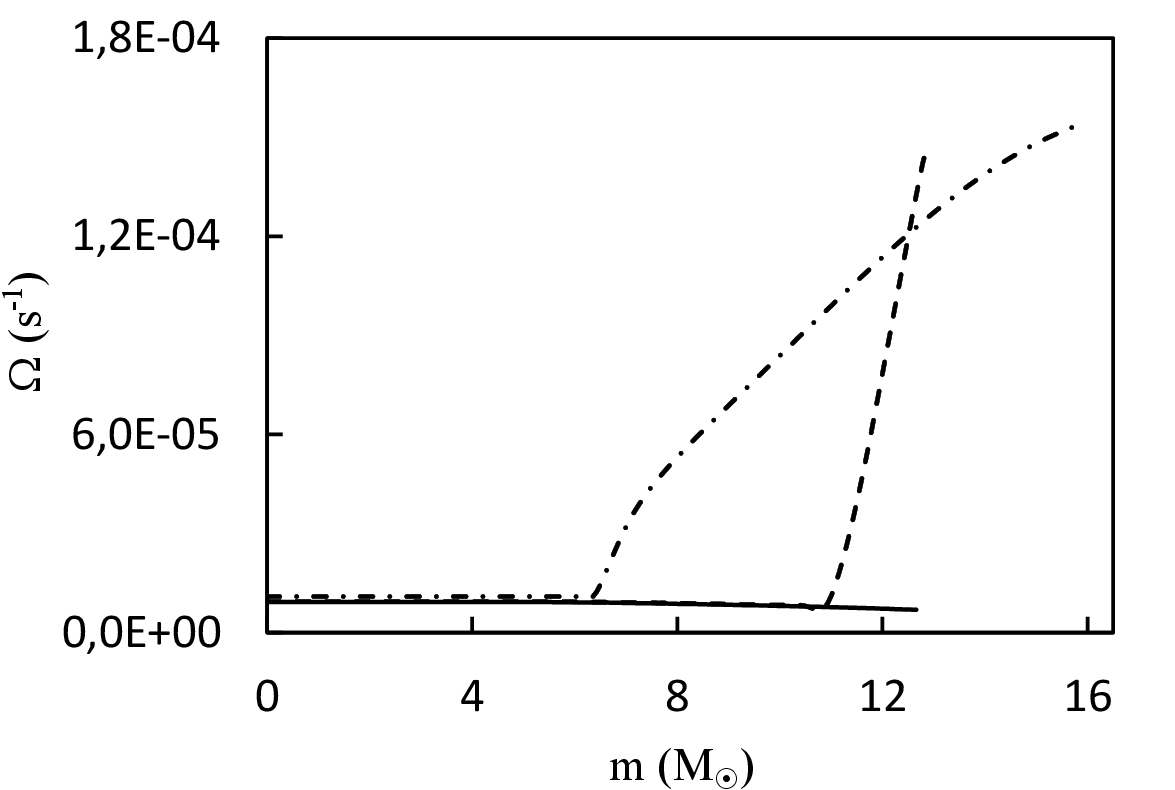}
\caption{Distribution of angular velocity inside the accretor with an initial mass of 12.8~$M_\odot$ (solid line) and after a mass increase of 0.24~$M_\odot$ (dashed line) and 3.2~$M_\odot$ (dashed-dotted line).}\label{add0102}
\end{figure}

\section{Calculation results}\label{sct3}
\subsection{Angular momentum transfer within and out of the accretor}\label{ssct31}

\begin{figure}[h]
\centering
\includegraphics[width=0.9\textwidth]{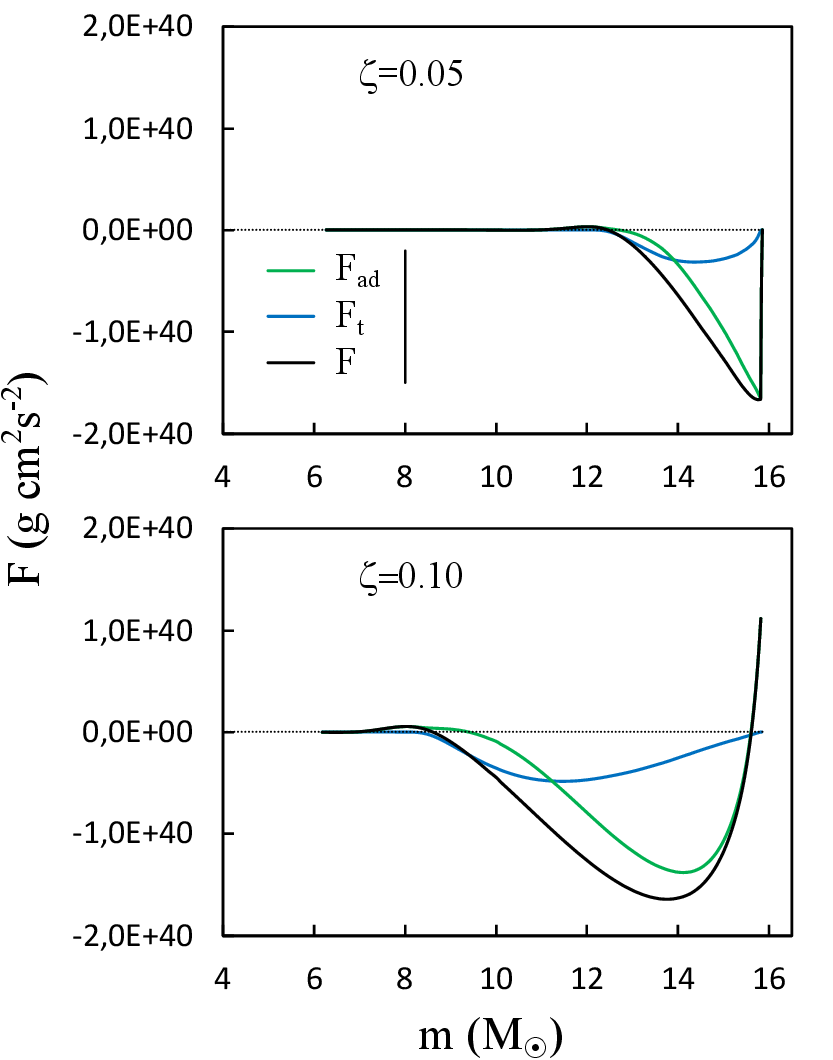}
\caption{The flux of angular momentum $F(m)$ (black line) in the interior of the accretor, as well as the advective flux $F_{ad}(m)$  (green line) and the turbulent flux $F_t(m)$  (blue line) at the end of mass transfer in the case of an inefficient boundary layer at $\zeta=0.05$ (top panel) and at $\zeta=0.10$ (bottom panel).}\label{fig02}
\end{figure}

\begin{figure}[h]
\centering
\includegraphics[width=0.9\textwidth]{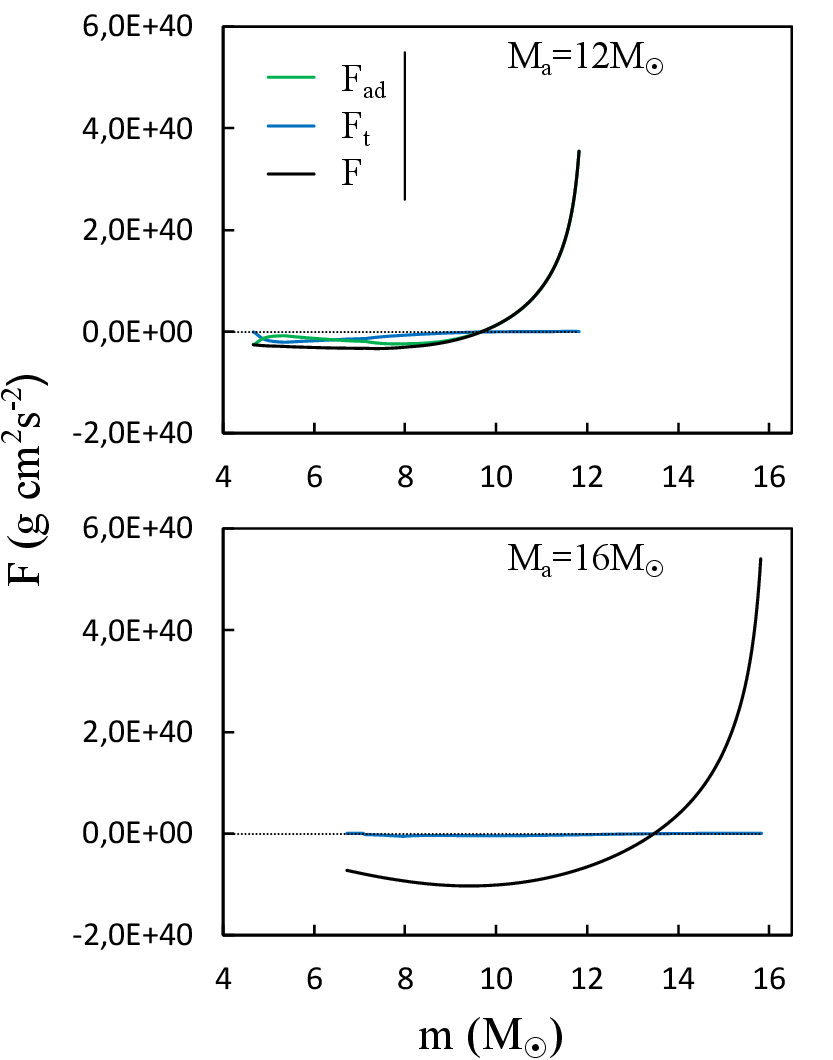}
\caption{Angular momentum flux $F(m)$ (black line) in the accretor interior, as well as advective flux $F_{ad}(m)$ (green line) and turbulent flux $F_t(m)$  (blue line) in the middle (top panel) and at the end (bottom panel) of mass transfer in the case of an effective boundary layer at $\zeta=0.50$.}\label{fig03}
\end{figure}

After the start of mass transfer in the binary system, the accreting component deviates from thermal equilibrium \citep{b70, fu77, nmns77}. The luminosity profile shown in Figure \ref{fig01} for an accretor with an initial mass of 8~$M_\odot$ is typical. Some of the released nuclear energy is spent to increase the entropy of the inner layers of the accreting component. The increase in luminosity in the upper layers occurs because of their contraction and the release of gravitational energy.

At the very beginning of the mass transfer process, a circulation cell forms in the accreted mass and in the underlying layer. The amplitude of the vertical component of the meridional circulation velocity shortly after the onset of accretion is shown in Figure \ref{add0101}. At the top of the accretor, the amplitude is positive. According to (\ref{eqU}), this means that at latitudes near the equator, the meridional flow is directed inwards the star, while at latitudes near the pole, the flow is directed outwards, and with this flow direction, the circulation transfers angular momentum inwards the star. In what follows, we will refer to this cell as 'the inner cell'. Below the inner cell, the amplitude of the vertical component of the meridional circulation velocity vanishes several times and changes sign. Where the amplitude is zero, the vertical component of the meridional circulation velocity is also zero. These locations mark the boundaries between circulation cells with different directions of angular momentum transfer. The circulation rate in this cascade of cells is much lower than in the inner cell. The circulation rate is lower the deeper the cell is located. Below the cascade of cells, the circulation rate is the same as it was before the onset of accretion and is of the order of $10^{-5}$~cm/s.

An example of the evolution of the angular velocity distribution is shown in Figure \ref{add0102}. The accreted layers rotate rapidly. As angular momentum is transferred inward, the underlying layers begin to rotate faster.

Circulation and turbulence within the inner cell transfer angular momentum toward the interior of the accreting component. The circulation rate is five orders of magnitude higher than in isolated stars. This significantly reduces the characteristic angular momentum transfer time in the accreting component compared to the isolated stars \citep{St19a}. The advective angular momentum flux in the accreting component is five orders of magnitude greater than in an isolated star. Circulation plays the largest role in transferring angular momentum in the upper part of the cell, whereas turbulence plays the largest role near the bottom (Figure \ref{fig02}). Due to turbulence, the angular momentum enters the layers below the cell. These layers begin to rotate more quickly and attach to the cell. The bottom of the cell then goes down into the accreting component.

Below the inner cell there is a cascade of circulation cells. 
The boundaries between cells descend into the star simultaneously with the lower boundary of the inner cell. Cells with opposite directions of angular momentum transfer successively span the same layer of the star.
The neighboring layers exchange a small amount of angular momentum as the cascade of cells passes through them into the deep inner layers of the accretor.


Due to accretion, the outermost layers of the accreting component begin to spin very quickly. Only when $\zeta=0.05$ and an inefficient boundary layer does the rotation of the accreting component remain slower than that of Kepler. In all other cases, the angular velocity of the surface of the accreting component reaches the critical value. This occurs after an increase in the mass of the accreting component by 2-4\% in the case of an effective boundary layer and by 4-7\% in the case of an ineffective boundary layer. After that, matter rotating at a critical velocity falls on the accreting component. In this matter, another circulation cell is formed. We will refer to this cell as 'the outer cell'. Circulation within the outer cell transfers some of the angular momentum from the accreted mass to the accretor surface. This part of the angular momentum enters the accretion disk and can be transferred further away from the accretor \citep{p1991, bk93}. The main process of angular momentum transfer in the outer cell is the meridional circulation. The turbulent transport in this cell is not significant (Figure \ref{fig02}, bottom panel). Due to the loss of angular momentum, the accreted layers contract. Their rotation velocity remains less than the critical one.

As the mass of the accreting component increases, the convective core also grows. The boundary of the convective core and the bottom of the inner cell move towards each other. At high values of $\zeta$, the bottom of the inner cell lowers to the convective core boundary before mass transfer ends. After that, the role of turbulent transport in the inner cell decreases. Circulation becomes the main process of angular momentum transfer in both the outer and inner cells (Figure \ref{fig03}). The role of turbulence is to lower the bottom of the inner cell deep into the accreting component.

\begin{figure}[h]
\centering
\includegraphics[width=0.9\textwidth]{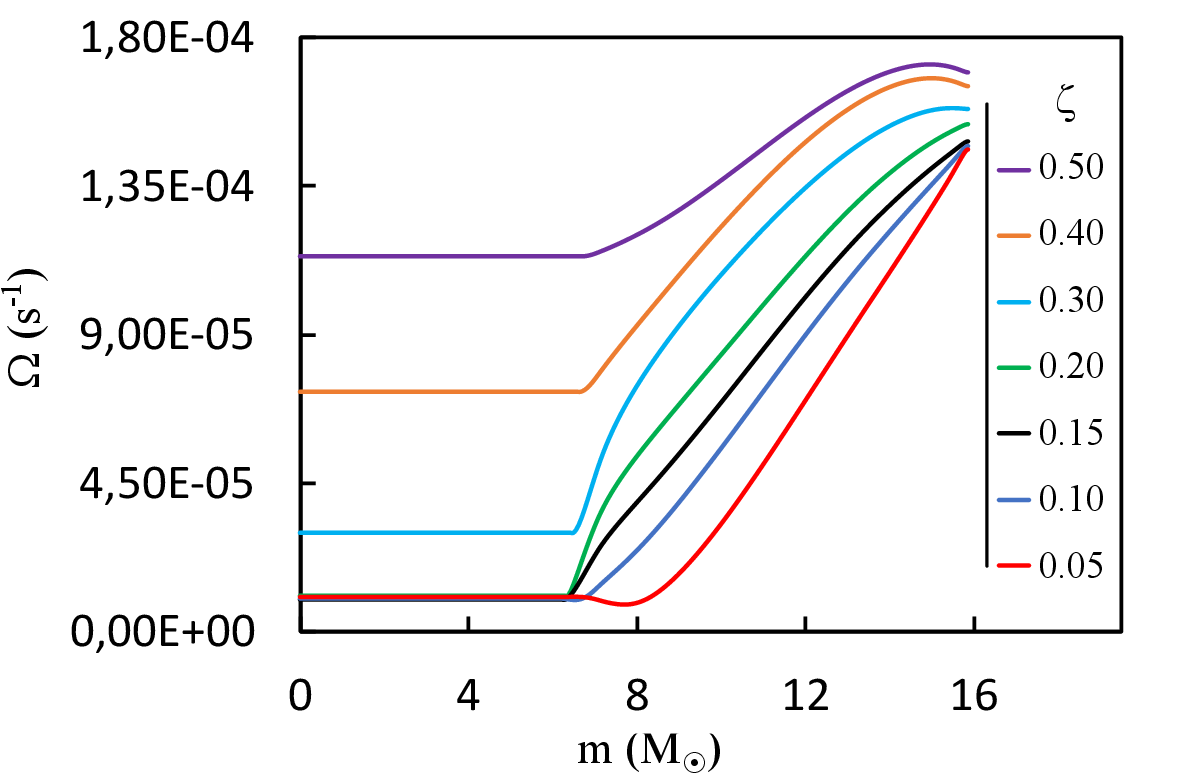}
\caption{Angular velocity in the depths of accretor after the end of the mass transfer in the case of effective boundary layer with $\zeta=0.05$ (red line), $\zeta=0.10$ (dark blue line), $\zeta=0.15$ (black line), $\zeta=0.20$ (green line), $\zeta=0.30$ (light blue line), $\zeta=0.40$ (orange line) and $\zeta=0.50$ (purple line).}\label{fig04}
\end{figure}

The distribution of angular velocity after the completion of mass transfer depends on the fraction of accreted mass $\zeta$ (Figure \ref{fig04}). Accreted layers and layers trapped by the inner circulation cell rotate more quickly. The bottom of the inner circulation cell is located deeper the greater the fraction of the accreted mass. At the end of the mass transfer process, the bottom of the inner cell at $\zeta<0.20$ is located within the radiative envelope of the accretor. The angular momentum of matter in the convective core is the same as before the mass transfer between the components begun. 

At $\zeta\ge0.20$, the bottom of the inner cell is lowered to the boundary of the convective core before the end of mass transfer. So, part of the angular momentum of the accreted mass entered the convective core during mass transfer. Additionally, convection also captures layers that have already increased their rotation due to the transfer of angular momentum. Therefore, at the end of the mass transfer, the convective core has increased angular momentum and faster rotation. 

In any case, the distribution of angular momentum in the accreting component after the completion of mass transfer is different from that in an isolated star with the same mass, angular momentum, and hydrogen content in the convective core. The outer layers of the accreting component rotate faster and the inner layers rotate more slowly than those of an isolated star (Figure \ref{fig04}). 

After completion of mass transfer, the circulation continues to transfer angular momentum from the outer layers of the accreting component to the inner layers as shown by \cite{St19a, St2021, St2022, St2024b}. During the thermal time scale, the angular velocities of the inner and outer layers become approximately the same.

\subsection{Lost and retained angular momentum}\label{ssct32}

\begin{figure}[h]
\centering
\includegraphics[width=0.9\textwidth]{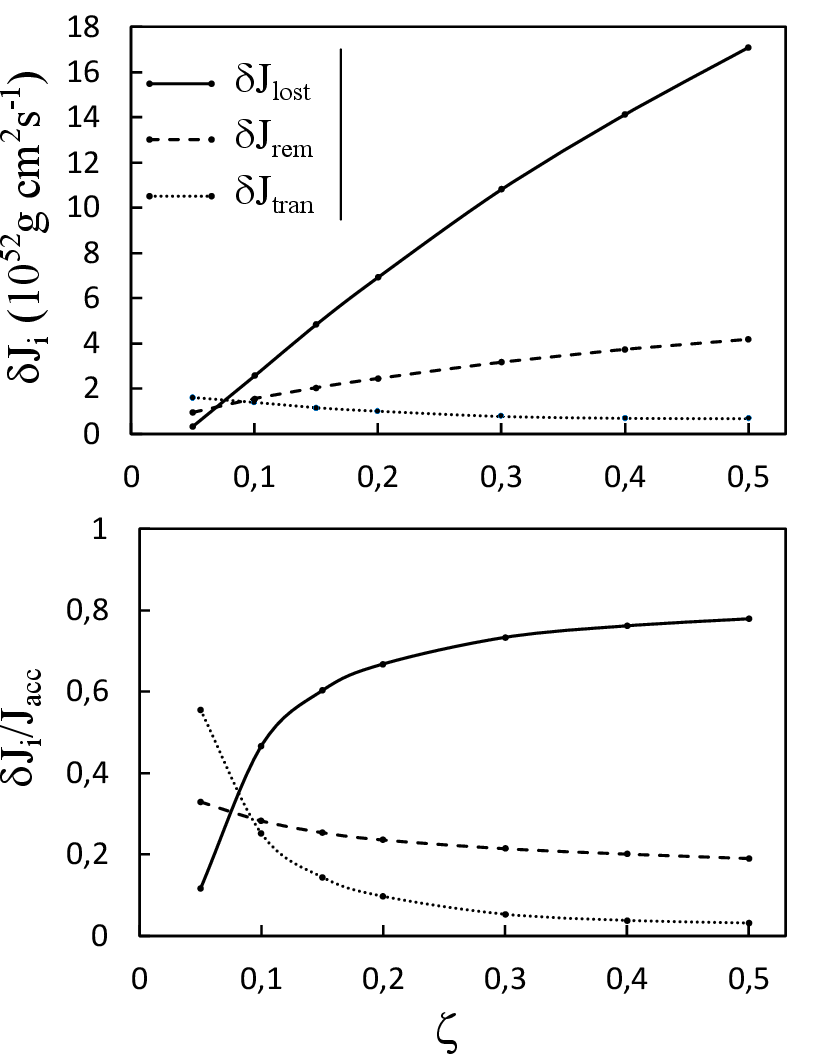}
\caption{The amount (top panel) and fraction (bottom panel) of the angular momentum transferred inside the accretor from the accreted mass $\delta J_{tran}$ (dotted line), remaining in the accreted mass $\delta J_{rem}$ (dashed line) and transferred to the surface of the accretor $\delta J_{lost}$ (solid line) by the end of mass transfer, depending on the fraction of the accreted mass $\zeta$.}\label{fig05}
\end{figure}

Let us denote the amount of angular momentum brought with the accreted mass during the mass transfer in a binary system as $J_{acc}$. This amount of angular momentum is divided into three parts: $\delta J_{lost}$ is the amount of angular momentum transferred from the accreted mass to the surface of the accretor and entered the disk, $\delta J_{rem}$ is the amount of angular momentum remaining in the accreted mass, $\delta J_{tran}$ is the amount of angular momentum transferred from the accreted mass to the inner layers of the accretor. Naturally, $J_{acc}= \delta J_{lost}+ \delta J_{rem}+ \delta J_{tran}$. Only in the case of an inefficient boundary layer at $\zeta=0.05$, the angular velocity of the accretor surface remains less than the Keplerian velocity. In this case, the accretor retains the acquired angular momentum. In all other cases, the angular velocity of the accretor surface increases to the critical velocity before the end of the mass transfer process. In these cases, the accretor loses part $\delta J_{lost}$ of the angular momentum brought with the accreted mass $J_{acc}$ (Figure \ref{fig05}, top panel).

The transfer of a part $\delta J_{lost}$ of the angular momentum from the accreted mass to the surface of the accretor and the loss of this part of the angular momentum by the accretor are the main processes that determine the rotation of the accretor at $\zeta>0.10$. By the end of mass transfer, the accretor loses $50\%-80\%$ of the angular momentum previously brought with the accreted mass (Figure \ref{fig05}, bottom panel). The amount of angular momentum retained in the accreted mass $\delta J_{rem}$ increases with $\zeta$ (Figure \ref{fig05}, top panel). This effect has a trivial explanation: the larger the accreted mass, the greater its angular momentum. The amount of angular momentum transferred from the accreted mass to the interior of the accretor $\delta J_{tran}$ depends weakly on $\zeta$ (Figure \ref{fig05}, top panel).

The angular momentum that the accreting component receives at the end of the mass transfer in the binary system is shown in Table \ref{tbl} (sixth and seventh columns for effective and ineffective boundary layers, respectively). Most of the angular momentum of the accretor at $\zeta>0.1$ is located in an accreted mass (Figure \ref{fig05}, top panel). Even in the case of $\zeta>0.20$, when the angular momentum is transferred from the accreted mass to the convective core, most of the accretor’s angular momentum is in the accreted mass.

The angular momentum of the accretor depends on the efficiency of the input of the angular momentum from the boundary layer only if $\zeta$ is small (Figure \ref{fig06}). 
As the fraction of accreted matter increases in the accretor final mass, the meaning of the boundary layer as a source of angular momentum decreases. For conservative mass transfer, the difference in the final angular momentum of the accretor in the cases of effective and inefficient boundary layers is ~1\%.

\section{Discussion}\label{sec4}

The lower bound of the rotation velocity $V_e/V_{Kep}$ of Be-type stars of the early subclasses is not precisely defined. According to \cite{c2005}, this boundary is in the range between 0.4 and 0.6. An isolated main-sequence star with a mass of 16~$M_\odot$ has a $V_e/V_{Kep}$ ratio greater than 0.6 (0.4) if the angular momentum is greater than $4.03\cdot10^{52}erg\cdot s$ ($2.95\cdot10^{52}erg\cdot s$) \citep{St07}.

\subsection{$\zeta$ high value example}\label{ssct41}

\begin{figure}[h]
\centering
\includegraphics[width=0.9\textwidth]{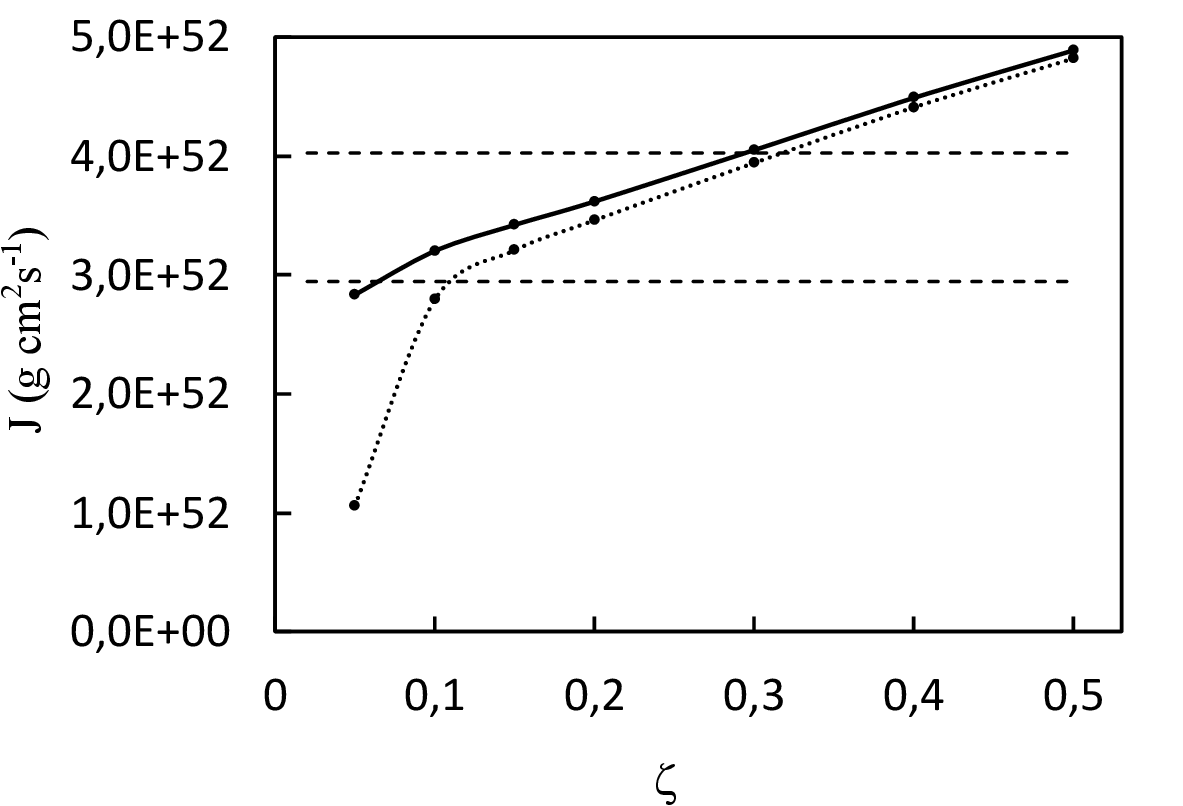}
\caption{The angular momentum of the accretor $J$ after mass transfer, depending on the accreted mass, in the case of an effective boundary layer (solid line) and in the case of an ineffective boundary layer (dotted line). The boundary value of the angular momentum above which the equatorial angular velocity of a main sequence star is always greater than 60\% (upper dashed line) and 40\% (lower dashed line) of the Keplerian value are also shown.}\label{fig06}
\end{figure}

The accretor receives an angular momentum greater than $4.03\cdot10^{52} erg\cdot s$ if, after the end of mass transfer, the fraction of accreted mass exceeds $30\%$ ($32\%$) in the case of an effective boundary layer (inefficient boundary layer) (Figure \ref{fig06}). With such an amount of accreted mass, the angular momentum of the accretor is practically independent of the efficiency of the transfer of angular momentum in the boundary layer. The angular momentum of the accretor is significantly less than the angular momentum of matter accreted from the Keplerian disk. The difference was transported by meridional circulation from the accreted matter to the surface of the star and entered the disk.

The angular momentum acquired by the accreting component as a result of mass transfer is only slightly dependent on the efficiency of the turbulence. At $\zeta$=0.33, the accreting component receives a rotation similar to that of early Be stars, even in the case of artificially suppressed turbulence \citep{St2022}. Some Algols possess sub-Keplerian disks \citep{kh82, 1984A&A...140..105C, k1988, k1989, r1992, rcf2014}. It cannot be ruled out that the same disks may be in more massive binary systems. However, the angular momentum of the accretor does not depend on the processes of decreasing the rotational velocity in the disk, if any. For example, a decrease in the disk rotational velocity by no more than 2.5 times does not affect the final value of the angular momentum of the accretor \citep{St2024a}.

Thus, at $\zeta>0.3$, the accreting component receives rotation like that of a early Be-star, regardless of how the angular momentum transfer processes in the boundary layer operate, what the turbulence efficiency is and how the disk rotates.

The accretor receives an angular momentum greater than $2.95\cdot10^{52} erg\cdot s$ if, after the end of mass transfer, the fraction of accreted mass exceeds $7\%$ ($12\%$) in the case of an effective boundary layer (inefficient boundary layer). The possibility of classifying the rotation of the accretor at $0.1\le\zeta<0.3$ as typical for early Be stars depends on the position of the lower limit of rotation of stars of this type.

Several of the observed properties of Be stars can be explained with large values of $\zeta$. The lower limit of the mass of Be stars in X-ray binaries can be explained if the Be star has accreted more than 30\% of the mass lost by the donor \citep{2020MNRAS.498.4705V}. The periods of Galactic Be binaries with a helium star companion and the masses of Be stars in these binaries can be explained by a conservative mass exchange in the Hertzsprung gap \citep{sl21}. The binary systems $\phi$ Per and V658 Car were formed as a result of conservative mass exchange. The $\phi$ Per system is one of the most studied systems with a Be-star \citep{gbfk98, sgmg18}. The V658 Car system is the only system with a Be star for which both a light curve and a radial velocity curve have been obtained \citep{2023BASBr..34..119D, 2025dath}. The rotation of Be stars in these systems can be predicted by calculating conservative mass exchange \cite{St2024b}.

\subsection{$\zeta$ low value example}\label{ssct42}

With a $\zeta$ of no more than $\sim0.1$, the accreting component obtains angular momentum near the  minimum possible lower limit of the rotation of the early Be-stars (Figure \ref{fig06}). Assuming an instantaneous redistribution of angular momentum to a state of rigid body rotation, the accreting component acquires a Keplerian break up rate after increasing its mass by a few percent \citep{p1981}. Meridional circulation and turbulence do not redistribute angular momentum instantly. The rate of lowering of the bottom of the inner circulation cell and, consequently, the amount of angular momentum transferred from the accreted mass to the interior of the accretor depends on the intensity of the turbulence \citep{St2022, St2024a}. Here we have considered the upper limit of the efficiency of turbulent transport. However, the angular velocity of the inner layers of the accretor remained lower than that of the accreted mass (Figure \ref{fig04}). After the transfer of angular momentum from the outer to inner layers of the accreting component, which follows the completion of the mass transfer in the binary system, the angular velocity of the surface of the accreting component will decrease to about 40\% of the Keplerian value and even lower.

At $\zeta\sim0.1$, the accretor can get a rotation like that of early Be stars during the common envelope stage \citep{St2021}. This case is characterized by high accretion rates. In the considering case of mass transfer, the accretion rate is one and a half orders of magnitude lower. In binary systems with a larger initial distance between the components, the thermal time scale of the donor is shorter. This may favor a higher accretion rate and, as a result, a faster downward movement of the bottom of the inner circulation cell. The initial synchronous rotation of the accreting component is slow. At the beginning of mass transfer in a binary system, the angular momentum of the accreted mass is also not very large. Therefore, after completion of the mass transfer, the angular momentum of the accreting component is different in the cases of effective and ineffective boundary layers. With a faster initial rotation, the difference between the cases of effective and ineffective boundary layers is smaller \citep{St19a}. In addition, the initial angular momentum of the accreting component is larger. A faster initial rotation may contribute to higher angular momentum values for the accreting component after completion of mass transfer. Therefore, the possibility of the formation of a early Be-star in the event of an increase in the mass of the accreting component by less than 10\% may depend on the initial parameters of the binary system. 


If the conditions of the formation and existence of the disk are such that the rotation velocity is less than the Keplerian one, then the angular momentum of the accreting component will again be smaller.

After a certain period of time following the end of the mass transfer, the rotation of the accretor becomes identical to the rotation of an isolated star with the same mass, angular momentum, and hydrogen content in the convective core. Most of this transformation occurs during the thermal time scale of the accretor immediately after the mass transfer, and then due to evolution, namely the contraction of the core and the expansion of the envelope \citep{St19a, St2021, St2022, St2024b}. If the tidal interaction is weak, the evolution of the accretor’s rotation is the same as that of an isolated star. The evolution of a rotating main-sequence isolated star is accompanied by the removal of angular momentum and helium from the inner layers to the outer ones \citep{mm05, St07, emmb08, geg13}. The intense removal of helium from the convective core into the radiative envelope activates the removal of angular momentum into the outer layers of the star. This lowers the boundary value of the angular momentum, above which the outer layers of the star show rapid rotation \citep{St09}. Unfortunately, the intensity of removal of helium from the convective core into the radiative envelope remains unclear \citep{St18}.

A synthesis of the X-ray Be star population, made on the assumption that only $\sim5\%$ of the mass lost by the donor falls on the accreting component for the formation of a Be star, predicts a large number of Be stars with masses of 3-5~$M_\odot$. This contradicts the observed lower limit of the mass of Be stars in X-ray binaries \citep{rkd2024}. Perhaps an increase in the mass of the accreting component by no more than $\sim10\%$ is not the main way of Be-stars forming.

The angular momentum of the accreting component can have different values when the gain in its mass does not exceed 10\%. These angular momentum values are determined by the initial parameters of the binary system, the angular momentum transfer mechanisms operating in the boundary layer, the efficiency of turbulence in the interior of the component, the angular velocity of the disk, and the position of the lower limit of rotation required for the formation of Be stars. The conditions under which the accreting component can acquire a rotation typical of early-type Be stars in the case of such a small mass gain require further research.

\section{Conclusion}\label{sec5}

We have considered the spinning up of the accreting component during mass transfer in a binary system in the Hertzprung gap as a function of the mass gain of the component. The mass of the accreting component after the end of the mass transfer is in all cases equal to 16~$M_\odot$, a value that is typical for Be stars of the early spectral subclass. The transfer of angular momentum in the interior of the component occurs due to meridional circulation and turbulence.  A special case is considered when, before mass transfer, the rotation of the accreting component is synchronous with the orbital motion.

The accreting component acquires critical rotation after increasing its mass by 2-7\%. Further mass growth occurs in a state of critical rotation of the accreting component. The main process that determines the rotation of the accreting component is the transfer of a large part of the angular momentum from the accreted mass to the surface of the accreting component due to meridional circulation. Then the angular momentum is transferred from the surface of the accreting component into the disk, as shown by \cite{p1991}, \cite{bk93} and \citep{mlv25}. The accreting component loses 50-80\% of the angular momentum received with the accreted mass if its mass gain exceeds 10\% by the end of mass transfer in the binary system. 

After the mass transfer is complete, most of the angular momentum of the accreting component is in the accreted mass. The angular momentum of the accreting component does not depend on how much angular momentum comes from the boundary layer if the increase in mass of the accreting component exceeds 10\%. The angular momentum of the accreting component was previously shown to be weakly dependent on the efficiency of turbulence in the interior of the component and does not depend on a possible decrease in disk rotation velocity below the Keplerian value \citep{St2022, St2024a}. The accreting component receives a rotation typical for stars of the Be type of early spectral subclass if the increase in its mass exceeds 30\%. This conclusion was obtained for the initial synchronous rotation of the accreting component and remains valid for faster initial rotation. Faster rotation of the component before mass transfer leads to faster rotation after mass transfer \citep{St19a}.

The possibility of classifying the rotation of the accreting component as typical for early Be-type stars, when the initial rotation is synchronous and the mass increase is from 10\% to 30\%, is determined by the exact position of the lower rotation boundary that allows the formation of the decretion disk. 
If the fraction of accreted mass does not exceed 10\%, the rotation of the accretor depends on a number of different factors.

\bmhead{Acknowledgements}
This work was carried out with the financial support of the Ministry of Science and Higher Education of the Russian Federation (theme FEUZ-2026-0012)

\bmhead{Author contributions} 
The author ES is the sole contributor to the preparation and writing of this work.

\bmhead{Ethics declaration} 
Not applicable.


\bibliography{2026_StApSS.bib}

\end{document}